 \documentclass[conference]{IEEEtran}
\ifx\pdfoutput\undefined
\bibliography{mybib}

\usepackage{graphicx}
\else
\usepackage[pdftex]{graphicx}
\fi
\usepackage{url}
\usepackage{array}
\usepackage{stfloats}
\usepackage{amssymb}
\usepackage{amsmath}
\usepackage{cite}
\usepackage{enumerate}
\usepackage{epstopdf}
\begin{document}
 \title{ Joint Channel Estimation and Pilot Allocation in Underlay Cognitive MISO Networks}
 \author{
  Maha Alodeh \quad Symeon Chatzinotas \quad Bj\"{o}rn Ottersten\\
  \authorblockA{
  Interdisciplinary Centre for Security, Reliability and Trust, University of Luxembourg\\
  4, rue Alphonse Weicker, L-2721 Luxembourg\\
  e-mail:\{maha.alodeh, symeon.chatzinotas, bjorn.ottersten\}@uni.lu
               }
 } 
 
 \maketitle
 \begin{abstract} 
Cognitive radios have been proposed as
agile technologies to boost the  spectrum utilization.
This paper tackles the problem of channel estimation and its impact on downlink
transmissions in an
underlay cognitive radio scenario. We consider
primary and cognitive base stations, each equipped with multiple
antennas and serving multiple users. Primary networks often suffer from the
cognitive interference, which can be mitigated by deploying beamforming at
the
cognitive systems to spatially direct the transmissions away from the primary
receivers. The accuracy of the estimated channel state information (CSI) plays an important role in designing accurate beamformers that can regulate the amount of interference. However, channel estimate is affected by interference.
Therefore, we propose different channel estimation and pilot allocation techniques to deal with the channel estimation at the cognitive systems, and to reduce the impact of contamination at the primary
and cognitive systems. 
 In an effort to tackle the contamination problem in primary and cognitive
 systems, we exploit the information
 embedded in the covariance matrices to successfully separate the channel estimate from other users' channels in correlated cognitive single input multiple input (SIMO) channels. A minimum mean
square error (MMSE) framework is proposed by utilizing the second order
statistics to separate the overlapping spatial paths that create the interference. We validate our algorithms by simulation and compare them to the state of the art techniques.
\let\thefootnote\relax\footnote{\textbf{978-1-4799-0959-9/14/\$31.00\copyright 2014
IEEE}}
\end{abstract}
%\begin{keywords}
{\bf Keywords:} Pilot contamination, pilot allocation, cognitive radio, channel estimation.
%\end{keywords}
\section{Introduction}
% 
% The cognitive radio has been proposed as an agile technology to tackle the spectrum scarcity and congestion contradiction. This agility is defined
% by the smart access of low priority users without violating the imposed constraints
% by the original users. As the fixed allocation techniques have assigned certain spectrum bands to certain applications, and these applications are no longer vastly used \cite{fcc}-\cite{federal}, this would require changing the regulations which is complicated and lengthy

 The paradigm of cognitive radio has been proposed as a promising agile technology that can revolutionize future of telecommunications by breaking the gridlock of the wireless spectrum \cite{goldsmith}-\cite{Haykin}. Two initial hierarchal levels have been defined: primary level and secondary level (the users within each level are called primary users (PU) and cognitive users (CU) respectively). Overlay,
underlay and interweave are three general techniques that can regulate the
coexistence terms of the two systems. The first two techniques permit simultaneous transmissions\cite{prior}-\cite{lee}, which leads to better spectrum utilization in comparison with the last one, which allocates the spectrum to the cognitive system by detecting the absence of the primary transmissions \cite{shree}.\smallskip

The use of multiple antennas at the primary and the cognitive base stations has proven to be very useful for the interference management in cellular
networks \cite{lee}. These characteristics make the multiple antenna techniques suitable to limit the
impact of the
created interference by cognitive transmissions on the primary receivers.
Therefore, the accuracy of the CSI has an important role on the interference
avoidance performance\cite{jose}.      
In this work, we focus on the CSI acquisition in an underlay cognitive network.  
The CSI acquisition in time division duplexing (TDD) systems has been handled in the literature by exploiting finite-length pilot sequences in the presence of cognitive interference.  Recently, the problem of non-orthogonality of training sequences
has been thoroughly investigated \cite{jose}-\cite{MAHA} in a multicell environment. It is pointed out in \cite{jose} that pilot contamination
degrades the performance and a robust precoding technique
is proposed to face this challenge. Specifically,
it is shown that the reuse of sequences across
interfering cells causes the interference mitigation performance
to rapidly degrade with the number of antennas, and thereby
undermines the benefits of MIMO systems in cellular networks.\smallskip

To allow the cognitive coexistence with the primary network, the interference at both
the estimation step and information transmission should be limited in order
not to degrade the primary system. A definition of the interference
constraint imposed by the primary system can be illustrated as follows   \smallskip  
\begin{itemize}
\item Contamination temperature $\mathcal{C}_{th}$: The amount of the interference
that can be tolerated by the primary base station (PBS) at the channel estimation phase.
\item Interference temperature $\mathcal{I}_{th}$: The amount of the cognitive downlink interference at the primary user (PU) receiver that can be accepted by the primary system\footnote{Interference
temperature $\mathcal{I}_{th}$ is usually defined for downlink transmissions
to design the beamforming at the cognitive system. This is out of scope
of this work and it is handled in \cite{lee}.}.
\end{itemize}
In this paper, we study the performance of the primary and cognitive networks
considering a pilot reuse between these two networks. We investigate
the impact of the pilot reuse on the accuracy of the estimation at the  primary
system. Moreover, we examine
the estimation procedure at the cognitive system and investigate the tradeoff
between
the multiuser diversity and the pilot contamination. Pilot
allocation
techniques at cognitive base station (CBS) are used to reduce the contamination
at the primary and cognitive systems which consequently have an impact on
the
downlink performance.  \smallskip

The adopted notations in the paper are as follows: we
use uppercase and lowercase boldface to denote matrices and vectors. Specifically, $\mathbf{I}_K$
denotes the $K\times K$ identity matrix. Let $\mathbf{X}^T$, $\mathbf{X}^*$ and
$\mathbf{X}^H$ denote the transpose, conjugate, and conjugate transpose
of a matrix $\mathbf{X}$ respectively. The Kronecker product of two matrices $\mathbf{X}$
and $\mathbf{Y}$ is denoted $\mathbf{X}\otimes\mathbf{Y}$.  Let $tr(\mathbf{X})$ denote the trace operation, and $\mathcal{CN}(a,\mathbf{R})$ is used to denote the circularly symmetric
complex Gaussian distribution, with the mean $a$ and
the covariance matrix $\mathbf{R}$.\smallskip
% \begin{figure}[h]
% \begin{center}
% \includegraphics[scale=0.63]{cr.eps}
% \caption{\label{PA}\textit{\small The normalized MSE vs number of antennas $M=20$,$P=0
% dB$, $\beta=1$,
% $\theta_1=\{10^{\circ},25^{\circ}\}$, $\theta_2=\{20^{\circ},35^{\circ}\}$,$\Delta\theta=20$,
% and $\Delta_o\theta=5^{\circ}$ }}\end{center}
% \end{figure}
\section{System model}
\label{sys}
Our model consists of primary 
cells with full spectrum reuse that coexist with a cognitive network. Estimation of flat block fading, narrow band
channels in the uplink is considered. The base station acquires the channel
estimate through uplink pilots transmitted by users. We assume that the pilot sequences, of length $\tau$ symbols, are used by single-antenna
users. All base stations
are equipped with an $M$-element uniform linear array (ULA) of antennas. It is assumed that each primary user is allocated an orthogonal pilot, so that no contamination occurs within the primary network. However, this pilot may be reused due to the limited resources by multiple cognitive users who contribute to the contamination of both primary and cognitive channel estimation. The pilot sequences used for estimating the user channels are denoted by $\mathbf{s}_i\triangleq\begin{array}{ccc}[s_{i1}&\hdots&s_{i\tau}]^T\end{array}\in\mathbb{C}^{1\times \tau}$. The pilot symbols are normalized such that $\{|s_{ij}|^2=\frac{P_t}{\tau},
\forall
j\in \tau\}$, where $P_t$ is the total pilot power. For the sake of simplicity,
we assume single PBS and CBS, where PUs use orthogonal pilots and these pilots
are reused to estimate the CU's channel with respect to CBS with the possibility
of reusing pilots within the cognitive systems.
The users' channel vectors are assumed to be $ \mathbb{C}^{M\times1}$
Rayleigh fading with correlation due to the finite multipath angle spread seen from the base station (BS) side. The channel between user $x$ and BS $z$ is denoted $\mathbf{h}_{xz}\sim\mathcal{CN}(0,\alpha_{xz}\mathbf{R}_{xz})\in
\mathbb{C}^{M\times 1}$, where $\alpha_{xz}$ is the attenuation from the
user $x$ to BS $z$. We denote
the channel covariance matrix $\mathbf{R}_{xz}\in\mathbb{C}^{M \times M}$.
We use the notation of $P$, $C$ for primary system and cognitive system elements
(i.e. BS or users) respectively. As multiple
CUs exist in the system, we use the index $j$ to distinguish the different
CUs.
Considering the transmission of $\mathbf{s}_i$
sequence, the $ M\times {\tau}$  signal baseband symbols sampled at the PBS
can be simplified as 
\vspace{-0.2cm}
\begin{eqnarray}
\mathbf{Y}_{P}=\mathbf{h}_{PP}\mathbf{s}^T_i+\displaystyle\sum_{\forall j\in \mathcal{K}_i}\mathbf{h}_{CP,j}\mathbf{s}^T_i+\mathbf{N}_{P},
\end{eqnarray}
where $\mathbf{h}_{PP}$ is the channel of interest at PBS. The sampled baseband signal at CBS
\vspace{-0.2cm}
\begin{eqnarray}
\mathbf{Y}_{C}=\mathbf{h}_{PC}\mathbf{s}^T_i+\displaystyle\sum_{\forall j\in \mathcal{K}_i}\mathbf{h}_{CC,j}\mathbf{s}^T_i+\mathbf{N}_{C}
\end{eqnarray}
%\vspace{-0.1cm}
where $\mathbf{h}_{SS,j}$ is the channel to be estimated at CBS. Moreover, $\mathcal{K}_i$ denotes the set of all CUs who use the training
sequence $\mathbf{s}_i$ simultaneously with the primary user. 
% For the sake of simplicity and without lack
% of generalization, we drop the pilot sequence
% index and assume that a single pilot is used over the network, which makes
% the baseband signal sampled at $c^{th}$ base station as
% \vspace{-0.3cm}
% \begin{equation}
% \label{y}
% \mathbf{Y}_c=\displaystyle\sum_{l=1}^{C} \mathbf{h}_{lc}\mathbf{s}^T+\mathbf{N}_c
% \end{equation}
$\mathbf{N}_S,\mathbf{N}_P\in\mathbb{C}^{M\times \tau}$ denotes the spatially and temporally white
complex additive Gaussian noise (AWGN) with element-wise variance $\sigma^2$
at CBS and PBS respectively. As we study the impact of reusing a single pilot
in the primary and cognitive system, the pilot indices can be dropped. 
Furthermore, we assume that the cognitive uplink transmissions are synchronized
with primary uplink transmissions. 
The contamination can occur in two cases:   
\begin{itemize}
\item The contamination is created at the estimation process at PBS due to
the reused pilots in the cognitive system.  
\item The contamination is created in the estimation process at CBS due to
the reused pilots in both cognitive and primary systems. \smallskip
\end{itemize}
\vspace{-0.2cm}
\subsection{Channel Model}
We consider a uniform linear array (ULA) at the BSs whose
response vector
can be expressed as
\begin{eqnarray}
\mathbf{a}(\omega)=\begin{array}{cccc}[1&e^{-j\omega}&\hdots&e^{-j(M-1)\omega}]^T\end{array}
\end{eqnarray}
where $\omega=\frac{2\pi d\sin\theta}{\lambda}$, $d$ is the antenna spacing at the base station,
$\lambda$ is the signal wavelength and $\theta$ is angle of arrival of a single
path. Assuming a flat fading channel, the received signal at the base station can be
expressed as a multipath model utilizing the response array vector as
 \vspace{-0.1cm}
 \begin{eqnarray}
 \label{multipath}
 \mathbf{h}=\sum_{i=1}^{Q}\gamma_i\mathbf{a}(\omega_i)
 \end{eqnarray}
 where $\gamma_{i}$ is a complex random gain factor, $\omega_i$ depends on
 the angle $\theta_i$ of the $i^{th}$ path, $Q$ is the number of paths.  
 A general correlation structure can be well approximated for limited angular
 spread by \cite{perz}
  \vspace{-0.05cm}
 \begin{eqnarray}
 \nonumber
 \mathbf{R}=\mathbf{D}_a\mathbf{B}^{\sigma_{\omega}}\mathbf{D}^H_{a}
 \end{eqnarray}
where $\sigma_{\omega}=2\pi\frac{d}{\lambda}\sigma_{\theta}\cos\theta$, $\mathbf{D}_a=\text{diag}[\mathbf{a}(\omega)]$.\, $\sigma_{\theta}$ is
the standard deviation of the angular spread. The matrix $\mathbf{B}^{\sigma_{\omega}}$ depends on the angular spread of the
multipath components. The angular distribution is Gaussian ${\omega}\in\mathcal{N}(0,\sigma^2_{\omega})$,
and it can be written as\smallskip
\vspace{-0.2cm}
 \begin{eqnarray}
 \label{gaussian}
 [\mathbf{B}^{\sigma_{\omega}}(m,n)]=e^{((m-n)\sqrt{3}\delta_{\omega})^2/2}.
 \end{eqnarray}
When ${\omega}$ is uniformly distributed over $[-\delta_{\omega},\delta_{\omega}]$,
 the covariance has the following structure
  \vspace{-0.05cm}
 \begin{eqnarray}
 \label{uniform}
 [\mathbf{B}^{\delta_{\omega}}(m,n)]=\frac{\sin((m-n)\delta_{\omega})}{(m-n)\delta_{\omega}}.
 \end{eqnarray}
and $\sigma_{\omega}=\sqrt{3}\delta_{\omega}$.\smallskip
\begin{newtheorem}{thm}{Theorem\cite{caire-s}}
\begin{thm}
The asymptotic normalized rank of the Toeplitz channel covariance matrix $\mathbf{R}$
with antenna separation $d$ and angle of arrival $\theta$ and angular
spread is given by
\begin{eqnarray}
\rho=\min\{1, B(d,\theta, \delta_{\omega})\}, 
\end{eqnarray}
where
$B(d,\theta, \delta_{\omega})=|d\sin(\theta-\delta_\omega)-d\sin(\theta+\delta_\omega)|.$
\end{thm}
\end{newtheorem}
From theorem 1, it can be noted that the rank of the user's covariance
is a function of the angular spread and direction of arrivals. The
users' positions with  respect to the surrounding BSs have a direct impact
on their channels, and as consequence the estimation procedures of these
channels. As a result, employing pilot allocation techniques that take into
the account the user's natural separability can boost the quality of estimation
at both PBS and CBS.      
\subsection{The CSI acquisition at the primary and cognitive systems}
The covariance information of
the target users and interfering users can be acquired   
exploiting resource blocks where the desired user and
interference users are known to be assigned pilot sequences
at different times. Alternatively, this information can be obtained using the
knowledge of the approximate users' positions and the type of the angular
spread at BS side exploiting the correlation equations (\ref{gaussian})-(\ref{uniform}). In this work, we assume two levels of covariance knowledge
\begin{itemize}
\item Coordinated knowledge, in which the PBS and CBS have covariance information
between themselves and the primary and cognitive users.  
\item  Cognitive knowledge, in which only the CBS has the covariance knowledge between
itself and all users in both systems.
\end{itemize}
Depending on correlation information availability on the CBS and PBS,
we propose different estimation and pilot allocation techniques in the following
sections.
 \vspace{-0.1cm}      
\section{Channel estimation for Underlay Cognitive Scenario}
\label{LS}
\vspace{-0.1cm} 
Utilizing the multiple antenna ULA structure, we propose a modified estimator with the target of decontaminating the reused pilots in the cognitive network. Our estimator exploits the information in the second order statistics of the channel vectors. The covariance matrices seize the required information of distribution (mainly mean and spread angle) of the multi-path signals at the base station \cite{kammeryer} and as shown in (\ref{gaussian}),(\ref{uniform}). We define a training matrix $\mathbf{S}=\mathbf{s}\otimes \mathbf{I}_M$, such that $\mathbf{S}^H\mathbf{S}=\tau\mathbf{I}_M$. Then, the received training
signal at the primary base station can be expressed as
\vspace{-0.1cm}
\begin{equation}
\label{rx}
\mathbf{y}_{P}=\mathbf{S}\Big(\mathbf{h}_{PP}+\displaystyle\sum_{\forall
j\in \mathcal{K}_i}\mathbf{h}_{SP,j}\Big)+\mathbf{n}_P
\end{equation}
where $\mathbf{n}_x\in \mathbb{C}^{M\tau\times 1}=\text{vec}(\mathbf{N}_x),
x\in\{S,P\}$ is the sampled noise at PBS. The sampled signal at CBS can be formulated as
\begin{equation}
\label{rx}
\mathbf{y}_{C}=\mathbf{S}\Big(\mathbf{h}_{PS}+\displaystyle\sum_{\forall
j\in \mathcal{K}_i}\mathbf{h}_{SS,j}\Big)+\mathbf{n}_S.
\end{equation}
\subsection{Naive Mean Square Error Estimation}
The estimator does not consider the interference at the estimation process
and can be formulated as
\begin{eqnarray}
\label{nmmse}
\mathbf{G}_n=\mathbf{R}_{PP}(\mathbf{R}_{PP}+\sigma^2\mathbf{I})^{-1}\mathbf{S}^H.
\end{eqnarray}

\subsection{Coordinated Minimum Mean Square Estimation}

The estimator at the PBS and CBS can be respectively expressed as
\begin{eqnarray}
\label{smmse}
\mathbf{G}_{P}=\mathbf{R}_{PP}\bigg(\mathbf{S}(\mathbf{R}_{PP}+\sum_i\mathbf{R}_{SP,i})\mathbf{S}^H+\frac{\sigma^2}{\tau}\mathbf{I}\bigg)^{-1},\\
\mathbf{G}_{C}=\mathbf{R}_{SS}\bigg(\mathbf{S}(\mathbf{R}_{SP}+\sum_i\mathbf{R}_{SS,i})\mathbf{S}^H+\frac{\sigma^2}{\tau}\mathbf{I}\bigg)^{-1}.
\end{eqnarray}
From (\ref{smmse}), it can be noted that the estimator at the PBS is a function
of all CUs' subspaces that utilize the same training sequence, which raises
the question about the possibility of acquiring the CUs' second order statistics
related to PBS. To reduce the impact of the contamination, the cognitive
system should have a pilot allocation strategy to reduce the contamination
on the primary system and cognitive system.\smallskip
\subsubsection{Mean Square Error Performance}
\vspace{-0.1cm}
% The minimum mean square error can be derived as 
% \vspace{-0.2cm}
% \begin{eqnarray}
% \label{mse}
% \eta=\mathbb{E}_{\mathbf{h}}\{tr\bigg(\hspace{-0.04cm}\big(\mathbf{h}_z-\mathbf{G}_x\mathbf{y}_x\big)\big(\hspace{-0.04cm}\mathbf{h}_z-\mathbf{G}_x\mathbf{y}_x\hspace{-0.08cm}\big)^H\bigg)\},
% \end{eqnarray}
 The estimation
errors at the PBS and CBS respectively can be expressed as 

\scriptsize
\begin{eqnarray}
\hspace{-0.5cm}\eta_{p}&=&tr\bigg(\mathbf{R}_{PP}-\mathbf{R}^2_{PP}\bigg(\mathbf{R}_{PP}+\sum_{\forall
j\in\mathcal{K}_i}\mathbf{R}_{SP,j}+\frac{\sigma^2}{\tau}\mathbf{I}\bigg)^{-1}\bigg)\\ \hspace{-0.5cm}\eta_{S}&=&\sum_{\forall j\in\mathcal{K}_i}tr\bigg(\mathbf{R}_{SS,j}-\mathbf{R}^2_{SS,j}\bigg(\mathbf{R}_{PS}+\sum_{\forall
l\in \mathcal{K}_i}\mathbf{R}_{SS,l}+\frac{\sigma^2}{\tau}\mathbf{I}\bigg)^{-1}\bigg).
\end{eqnarray}
\normalsize
\vspace{-0.12cm}
 From the previous equations, it can be concluded that the mean square errors
 at PBS and CBS are functions of the subspaces of the cognitive interfering
 users.
% can be limited as
% % \begin{eqnarray}
% % \eta_p&=&tr(\mathbf{R}_{P,P})\bigg(\alpha_{P,P}-\frac{(\alpha_{P,P})^2}{(\alpha_{P,P}+\sum_i\alpha_{S,P,i})}\bigg)
% % %&=&tr(\mathbf{R}_{P,P})\bigg(1-\frac{1}{C}\bigg)
% % \end{eqnarray}
% The same thing holds for the secondary systems $\eta_{S,i}$
\vspace{-0.15cm}
\subsection{Primary  Cognitive MMSE Estimator}
To minimize the MSE at the PBS, the contamination constraint should be taken
into consideration. The mean square error can be formulated as
\vspace{-0.2cm}
\begin{eqnarray}\nonumber
\mathcal{E}_p&=&\mathbb{E}\{\Bigg(\mathbf{h}_{PP}-\gamma^{-1}\mathbf{G}_P\mathbf{S}\big(\sum_{\forall
j\in \mathcal{K}_i}\mathbf{h}_{SP,j}+\mathbf{h}_{P,P}+\mathbf{n}_p\big)\Bigg)\\
&\times&\Bigg(\mathbf{h}_{PP}-\gamma^{-1}\mathbf{G}_P\mathbf{S}(\sum_{\forall
j\in\mathcal{K}_i}\mathbf{h}_{SP,j}+\mathbf{h}_{PP}+\mathbf{n}_p)\Bigg)^H\}.
\end{eqnarray}
The optimization problem that takes into the account the contamination effect can be formulated as
\vspace{-0.05cm}
\begin{eqnarray}
\begin{array}{cc}\underset{\mathbf{G}_P,\gamma}{\min}&\mathcal{E}_p\\
\text{s.t.}&\quad tr(\mathbf{G}_P\mathbf{G}^H_P)\leq P\\
\quad&tr(\mathbf{G}_P\mathbf{S}\sum_{\forall j\in \mathcal{K}_i}\mathbf{R}_{SP,j}\mathbf{S}^H\mathbf{G}^H_P)\leq \mathcal{C}_{th}.\\
\end{array}
\end{eqnarray}
\vspace{-0.1cm}
To solve the previous optimization problem, we need to express the associated Lagrange equation as 
\vspace{-0.1cm}
\begin{eqnarray}\nonumber
\hspace{-0.1cm}\mathcal{L}(\mathbf{G}_P)&\hspace{-0.1cm}=\hspace{-0.1cm}&tr\bigg(\mathbf{R}_{PP}-\gamma^{-1}\mathbf{G}_P\mathbf{S}\mathbf{R}_{PP}-\gamma^{-1}\mathbf{R}_{PP}\mathbf{S}^H\mathbf{G}^H_P\\\nonumber
&\hspace{-0.1cm}+&\hspace{-0.1cm}\gamma^{-2}\mathbf{G}\mathbf{S}(\sum_{\forall
j\in\mathcal{K}_i}\mathbf{R}_{SP,j}+\mathbf{R}_{PP}+\sigma\mathbf{I})\mathbf{S}^H\mathbf{G}^H_P\hspace{-0.1cm}\bigg)\\\nonumber
&\hspace{-0.1cm}+&\hspace{-0.1cm}\lambda(tr(\hspace{-0.04cm}\mathbf{G}_P\mathbf{G}^H_P)-P)\\\
&\hspace{-0.1cm}+&\mu\bigg(\hspace{-0.1cm}tr(\mathbf{G}_P\mathbf{S}\sum_{\forall
j\in\mathcal{K}_i}\mathbf{R}_{SP,j}\mathbf{S}^H\mathbf{G}^H_P)-\mathcal{C}_{th}\hspace{-0.1cm}\bigg)
\end{eqnarray}  
where
$\gamma$
indicates a scaling factor for the received signal. The corresponding Karush-Kuhn-Tucker (KKT) conditions for $\mathcal{L}(\mathbf{G})$ can be written as
\begin{eqnarray}\nonumber
\hspace{-0.5cm}\label{1}
&-&\gamma^{-1}\mathbf{R}_{PP}\mathbf{S}^H+\gamma^{-2}\mathbf{G}_P\mathbf{S}\sum_{\forall
j\in\mathcal{K}_i}(\mathbf{R}_{SP,j}+\mathbf{R}_{PP})\mathbf{S}^H\\
&+&\lambda\mathbf{G}_P+\mu\mathbf{G}_P\mathbf{S}\sum_{\forall j\in \mathcal{K}_i}\mathbf{R}_{SP,j}\mathbf{S}^H=\mathbf{0},\\\nonumber
\label{2}
&+&\gamma^{-2}tr(\mathbf{G}_P\mathbf{R}_{PP}\mathbf{S}+\gamma^{-2}\mathbf{R}_{PP}\mathbf{S}^H\mathbf{G}^H_P\\\nonumber
&-&2\gamma^{-3}\mathbf{G}_P\mathbf{S}(\sum_{\forall j\in \mathcal{K}_i}\mathbf{R}_{SP,j}+\mathbf{R}_{P,P}+\sigma\mathbf{I})\mathbf{S}^H\mathbf{G}^H_P)=0,\\
\label{3}
&\lambda&\geq 0, tr(\mathbf{G}_P\mathbf{G}^H_P)-P\leq 0,\\
\label{4} 
&\mu&\geq 0,tr(\mathbf{G}_P\mathbf{S}\sum_{\forall j\in\mathcal{K}_i}\mathbf{R}_{SP,j}\mathbf{S}^H\mathbf{G}^H_P)-\mathcal{C}_{th}\leq
0,
\end{eqnarray}
\begin{eqnarray}
\label{5}
&\quad&\lambda \bigg(tr(\mathbf{G}_P\mathbf{G}^H_P)-P\bigg)=0,\\
\label{6}
&\quad&\mu \bigg(tr(\mathbf{G}_P\mathbf{S}\sum_{\forall j\in\mathcal{K}_i}\mathbf{R}_{SP,j}\mathbf{S}^H\mathbf{G}^H_P)-\mathcal{C}_{th}\bigg)=0.
\end{eqnarray}
From (\ref{1}), we can formulate the modified MSE estimator as follows
\begin{eqnarray}
\mathbf{G}_P=\gamma\mathbf{R}_{PP}\Big(\mathbf{R}_{PP}+\zeta_1\sum_{\forall
j\in\mathcal{K}_i}\mathbf{R}_{SP,j}+\zeta_2\mathbf{I}\Big)^{-1}\mathbf{S}^H
\end{eqnarray}
where $\zeta_1=\gamma\mu$, $\zeta_2=\gamma\lambda$, $\zeta_2=\frac{M}{\tau P}-\frac{2\mathcal{C}_{th}\zeta_1}{\sigma^2\tau
P}$. 
The final estimator can be expressed as
\vspace{-0.2cm}
\footnotesize
\begin{eqnarray}
\label{cmmse}
\hspace{-0.5cm}\mathbf{G}_P=\gamma\mathbf{R}_{PP}\Bigg(\mathbf{R}_{PP}+\zeta_1\sum_{\forall
j\in \mathcal{K}_i}\mathbf{R}_{SP,j}+\bigg(\frac{M\sigma^2-2\mathcal{C}_{th}\zeta_1}{\sigma^2\tau P}\bigg)\mathbf{I}\Bigg)^{-1}\mathbf{S}^H.
\end{eqnarray}\normalsize
To determine the values of $\zeta_1$, we need to define the following function
\begin{eqnarray}
f(\mu)=tr(\{P\sum_{\forall j\in \mathcal{K}_i}\mathbf{R}_{SP,j}-\mathcal{C}_{th}\mathbf{I}\}\mathbf{G}_P(\mu)\mathbf{G}^H_P(\mu)).
\end{eqnarray}
In order to ensure that the contamination does not exceed the threshold, this condition
should be considered $f(\mu)\leq0$. This condition results in $\frac{\mathcal{C}_{th}}{tr(\mathbf{G}_P(\mu)\mathbf{S}\sum_j\mathbf{R}_{SP,j}\mathbf{S}^H\mathbf{G}^H_P(\mu))}>\frac{P}{tr(\mathbf{G}_P(\mu)\mathbf{G}^H_P(\mu))}$
which makes $\gamma=\sqrt{\frac{P}{tr(\mathbf{G}_P(\mu)\mathbf{G}^H_P(\mu))}}$.
The value of $\zeta^*$ can be evaluated at CBS and passed to PBS as the knowledge
of second order statistics is available at CBS.   
\smallskip

\subsubsection{Mean Square Error Performance}
The contamination temperature can be  translated into mean square error constraint.
The MSE can be evaluated using (\ref{cmmse}), and has the following formulation:
\begin{eqnarray}
\eta_p=tr\Big(\mathbf{R}_{PP}-\mathbf{R}^2_{PP}\big(\mathbf{R}_{PP}+\zeta^*\sum_{\forall
j\in
\mathcal{K}_i}\mathbf{R}_{SP,j}+\mathbf{I}\big)^{-1}\Big).
\end{eqnarray}
It can be noted that MSE is a function of the contamination temperature.
By increasing the contamination temperature, the MMSE estimator reduces to
the same formulation as the typical MMSE estimator.
\vspace{-0.2cm}  
\section{ pilot decontamination using pilot allocation}
\vspace{-0.1cm}
To enhance the quality of estimation, we introduce pilot allocation algorithms
to assign the pilot to the set of the secondary users that span distinct
subspaces with respect to the PU and the set of cognitive user. Moreover,
this pilot allocation can simplify the estimation at the PBS by assigning the
same training sequence to a suitable set of CUs in the cognitive networks.
\vspace{-0.1cm}
\subsection{Optimal Pilot Allocation}
To find the optimal pilot allocation that achieves the minimum MSE across
the networks, we need to exhaustively search all possible combinations. To
simplify the search, we proposed low complexity greedy algorithms to find
the suboptimal set of cognitive users that can simultaneously utilize the
same training sequence with the primary users. These algorithms can be summarized
as follows   
\subsection{Greedy MSE Minimizing Pilot Allocation Algorithm}
We adopt the MSE as a metric to optimize the pilot allocation algorithm.  Define the set of the CUs that utilizes the training sequence
$\mathbf{s}$ as
$\mathcal{U}_s$, and the set of CUs that \textit{may} allocate the same pilot
with PU $\mathcal{U}$. 
Define the mean square error metric as follows

%\vspace{-0.15cm}
\footnotesize
\begin{eqnarray}
%\begin{array}{ccc}
\hspace{-0.8cm}\mathcal{\eta}_p(P,\mathcal{U}_s)&=&\mathbf{R}_{PP}\big(\mathbf{R}_{PP}+\sum_{i\in\mathcal{U}_s}\mathbf{R}_{SP,i}+\frac{\sigma^2}{\tau}\mathbf{I})^{-1},\\
\hspace{-1cm}\mathcal{\eta}_{s}(P,\mathcal{U}_s)&=&\hspace{-0.2cm}\sum_{j\in\mathcal{U}_s}\hspace{-0.1cm}\mathbf{R}_{SS,j}(\mathbf{R}_{PS}+\hspace{-0.2cm}\sum_{i\in\mathcal{U}_s\subset{\mathcal{U}}}\hspace{-0.2cm}\mathbf{R}_{SS,i}+\hspace{-0.1cm}\frac{\sigma^2}{\tau}\mathbf{I})^{-1}.
%\end{array}
\end{eqnarray}

\normalsize
It should be noted that these pilot allocation algorithms are designed  at
cognitive system deployment, so they are functions of the relative positions of
the cognitive users and primary user. 
\vspace{-0.25cm}
\footnotesize
\begin{center}
\begin{tabular}{p{8.3cm}}\\
\hline
\hline
\textbf{A.1}\quad Greedy Pilot Allocation Algorithm\\
\hline
\begin{itemize}
\item To reduce the pilot contamination at PBS  
\begin{enumerate}
\item Initialize the set of CUs that may allocate the same training sequence
with the $i^{th}$  PU $\mathcal{U}(\mathbf{s}_i)=\mathcal{\phi}$. 
\item If the PU allocates the training sequence of $\mathbf{s}_i$, 
$\underset{j^*\in \mathcal{C}}{\arg\min}\quad\eta_p(P,\mathcal{U}\cup \{j\})s.t.$
$\mathcal{U}\leftarrow\mathcal{U}\cup \{j^*\}$.
\item if $\eta_p<\zeta_{th}$, go to step 1.
\end{enumerate}
\item To reduce the pilot contamination within the cognitive system
\begin{enumerate}
\item Initialize $\mathcal{U}_s=\mathcal{\phi}$
\item $k^*=\arg\underset{k\in\mathcal{U}}{\min}\quad\eta_s(P,\mathcal{U}_s\cup
\{k\})$,
$\mathcal{U}_s\leftarrow\mathcal{U}_s\cup \{k^*\}$.
\end{enumerate}
\end{itemize}\\
\hline
\end{tabular}
\end{center}\smallskip
\normalsize
It can be noted that if the cognitive users are located in distinct position, they span different subspaces which can reduce the probability of having
 contamination in the primary and cognitive estimate. Therefore, this has a direct
 impact on the interference avoidance based technique in the downlink transmissions.
 \vspace{-0.25cm}
\subsection{Heuristic Pilot Allocation }
\vspace{-0.2cm}
Another pilot allocation that can handle a generic estimation technique is to assign the pilot based on the users spatial separability. We propose a new metric to express the amount of overlap in subspaces
\vspace{-0.1cm} 
\begin{eqnarray}
\label{sps}
\delta^p_{SP,i}=\frac{tr\big(\mathbf{R}_{PP}\mathbf{R}_{SP,i}\big)}{tr(\mathbf{R}_{PP})tr(\mathbf{R}_{SP,i})}
\end{eqnarray}
where $0\leq\delta^p_{SP,i}\leq 1$. When $\delta^p_{S,P,i}$ is close to 1, the
users span highly overlapped subspaces, but when $\delta^p_{S,P,i}$ is close
to 0, the users span a highly separated subspaces. To express the concatenated
subspaces of the CUs
overlapping with a PU, we define the following metric
\vspace{-0.1cm}
\begin{eqnarray}
\label{spss}
\delta^p_{SP}=\frac{tr\big(\mathbf{R}_{PP}\sum^C_{i=1,i\neq l}\mathbf{R}_{SP,i}\big)}{tr(\mathbf{R}_{PP})tr(\sum^C_{i=1,i\neq
l}
\mathbf{R}_{SP,i})}.
\end{eqnarray}\smallskip
If we define the semi-orthogonality threshold values between the primary
system and CUs as $\delta_p$, and between the CUs  $\delta_s$, The pilot allocation algorithm can be written as  
\footnotesize
\begin{center}
\begin{tabular}{p{8.3cm}}
\hline\hline
\textbf{A.2}$\quad\delta_p$, $\delta_s$ Heuristic Pilot Allocation\\
\hline 
\begin{itemize}
\item $\delta_p$ step to reduce the contamination at PBS
\begin{enumerate}
\item  Initialize the set of CU that may allocate the same training sequence
with the $i^{th}$  PU $\mathcal{U}(\mathbf{s}_i)=\phi$.
\item $\forall j\in \mathcal{C},\delta^p_{SP,j}\leq\delta_p$, $\mathcal{U}(\mathbf{s}_i)\leftarrow\mathcal{U}(\mathbf{s}_i)\cup
j$. 

\end{enumerate}
\item $\delta_s$ step to reduce the contamination at CBS
\begin{enumerate}
 \item Initialize $\mathcal{U}_s=\phi$
\item $k^*=\arg\underset{k\in\mathcal{U}}{\min}\quad\delta_s(P,\mathcal{U}_s\cup
k)$,
$\mathcal{U}_s\leftarrow\mathcal{U}_s\cup \{k^*\}$.
\end{enumerate}
\end{itemize}\\
\hline
\end{tabular}
\end{center}
\normalsize
\subsection{User Grouping Based Pilot Allocation} 
User grouping has been proposed in \cite{caire-g} for the purpose of utilizing
the users' correlation matrices to virtually sectorize the BS based on users'
channel statistics due to the rank limitation stated by Theorem (1).
To simplify the pilot allocation and the estimation at the both systems,
we cluster the PUs and CUs into different groups such that each user should
belong to two different groups. The first set of groups $\mathcal{G}_{P,g}$ is related to PBS  and the
other one $\mathcal{G}_{C,g}$ is related to CBS. These groups are designed
according to these guidelines
\begin{itemize}
\item The cognitive users in the same group should have channel covariance eigenspace
spanning a common subspace, which identifies the group.
\item The subspaces of the group should span mutually orthogonal subspaces
or disjoint ones (i.e. the groups have non-overlapping ). The $\mathcal{G}_{P,g}:\mathcal{G}_{P,g}\subset\mathcal{G}_{P,g},\displaystyle\cap_g\mathcal{G}_{P,g}=\phi,\cup_g\mathcal{G}_{P,g}=\mathcal{G}_P$,
$\mathcal{G}_{C,g}:\mathcal{G}_{C,g}\subset\mathcal{G}_C,\displaystyle\cap_g\mathcal{G}^i_{C,g}=\phi$.

\item The CUs distribution among PBS groups has no relation to their distribution
among CBS ones.
\end{itemize}\smallskip
 These factors depend on the users' relative positions
with respect to BSs (PBS, CBS) and the local scattering environment.
We use the chordal distance as a metric to assess the similarity among the
users, which makes it suitable for users grouping. Given two matrices $\mathbf{X}\in \mathbb{C}^{M\times r}$, $\mathbf{Y}\in \mathbb{C}^{M\times r}$ their chordal distance denoted by $d_c(\mathbf{X},\mathbf{Y})$
is defined by
\vspace{-0.2cm}
\begin{eqnarray}
d_c(\mathbf{X},\mathbf{Y})=\|\mathbf{X}\mathbf{X}^H-\mathbf{Y}\mathbf{Y}^H\|^2_F.
\end{eqnarray}
The group subspaces for the CUs are defined as $\mathbf{Q}_{ SP,g},\mathbf{Q}_{SS,g}\in \mathbb{C}^{M\times r}:g=1,\hdots,\mathcal{ G}$ are assumed
to be known and fixed a priori based on users' geometric distribution where
$r$ defines the rank of $\mathbf{Q}_{SS,g},\mathbf{Q}_{SP,g}$ , and
$\mathbf{U}_{k,x}$ is the $k^{th}$ users dominant eigenvectors. Assuming we have $\mathcal{M}$
groups, we can group the users
using the following algorithm 
\begin{itemize}
\item Select $x=SP$ or $x=SS$
\item for $g=1,\hdots, \mathcal{G}$, set $\mathcal{G}_{x,g}=\mathcal{\phi}$
\item for $m=1,\hdots, \mathcal{M}$
\begin{eqnarray}
d_c(\mathbf{U}_{x,m},\mathbf{Q}_{x,g})=\|\mathbf{U}_{x,m}\mathbf{U}^H_{x,m}-\mathbf{Q}_{x,g}\mathbf{Q}^H_{x,g}\|^2_F.
\end{eqnarray}
Find the minimum distance
\begin{eqnarray}
g=\arg\underset{g}{\min}\quad d_c(\mathbf{U}_{x,m},\mathbf{Q}_{x,g}),
\end{eqnarray}
and add user $k$ to group $g$, $\mathcal{G}_{x,g}=\mathcal{G}_{x,g}\cup k$.\smallskip
\end{itemize}
It is obvious that the performance depends on the selection of the predefined
subspaces $\mathbf{Q}_{x,g}$. 
\subsubsection{$\mathbf{Q}_{x,g}$ determination}
The set of $\{\mathbf{Q}_{x,g}\}, g=1, \hdots, \mathcal{G}_x$ are chosen to span disjoint
subspaces by assuming distinct angular spread
or have a minimal overlap
with
the other group which can be found using the chordal distance metric as follows
\begin{eqnarray}
d(\mathbf{\hat{Q}}_{x,g},\mathbf{\hat{Q}}^H_{x,j})=\arg\min \|\mathbf{{Q}}_{x,j}\mathbf{{Q}}^H_{x,j}-\mathbf{\hat{Q}}_{x,g}\mathbf{\hat{Q}}^H_{x,g}\|^2_F.
\end{eqnarray}
The user grouping is performed once for fixed users position. Based on the grouping,
we propose a new pilot allocation algorithm\\
\footnotesize
\begin{tabular}{p{8cm}}
\hline\hline
\textbf{A.3}\quad Group Based Pilot Allocation\\
\hline
\begin{itemize}
\item PBS selects the $j^{th}$  PU.
\item Acquiring this information, CBS finds the group  $\mathbf{Q}_{S,P,j\in
g^*}$ that falls within the selected PU subspace.
\item Select the $k\in\mathbf{Q}_{SP,g}\forall g\not\in g^*$
\item Find the subspace that has the minimum chordal distance $l^*=\arg\underset{l }{\min}\quad d(\mathbf{Q}_{SP,l},\mathbf{Q}_{PP,j\in
g^*})$.
\end{itemize}\\
\hline
\end{tabular}\\
%\end{table}
\normalsize
\smallskip
The user grouping pilot allocation can be combined with any of the described estimation techniques NMMSE, MMSE and CMMSE. 
\section{Numerical Results}
In order to assess the performance of the proposed
schemes, simulations of cognitive and primary systems have been performed.
The assumed scenario: single PU, $20$ SUs, $10$ antennas at CBS and PBS, the angular
spread is assumed to be $30^\circ$ uniformly distributed at CBS ULA with overlap of $6^\circ$. These
parameters are applied in the following simulations unless
otherwise stated. The users channels are assumed to have the formulation of (\ref{multipath}),
and undergo the correlation (\ref{gaussian}), (\ref{uniform}). The studied
metric is the normalized sum mean square error, which can be expressed for
PBS and CBS respectively as
\begin{eqnarray}
\eta_P&=&10\log_{10}(\frac{\|\mathbf{\hat{h}}_{PP}-\mathbf{h}_{PP}\|^2}{\|\mathbf{h}_{PP}\|})\\
\eta_C&=&10\log_{10}(\frac{\sum_{j}\|\mathbf{\hat{h}}_{SS,j}-\mathbf{h}_{SS,j}\|^2}{\sum_j\|\mathbf{h}_{SS,j}\|}).
\end{eqnarray} 
\vspace{-0.5cm}

\begin{figure}[h]
\begin{center}
\includegraphics[scale=0.48]{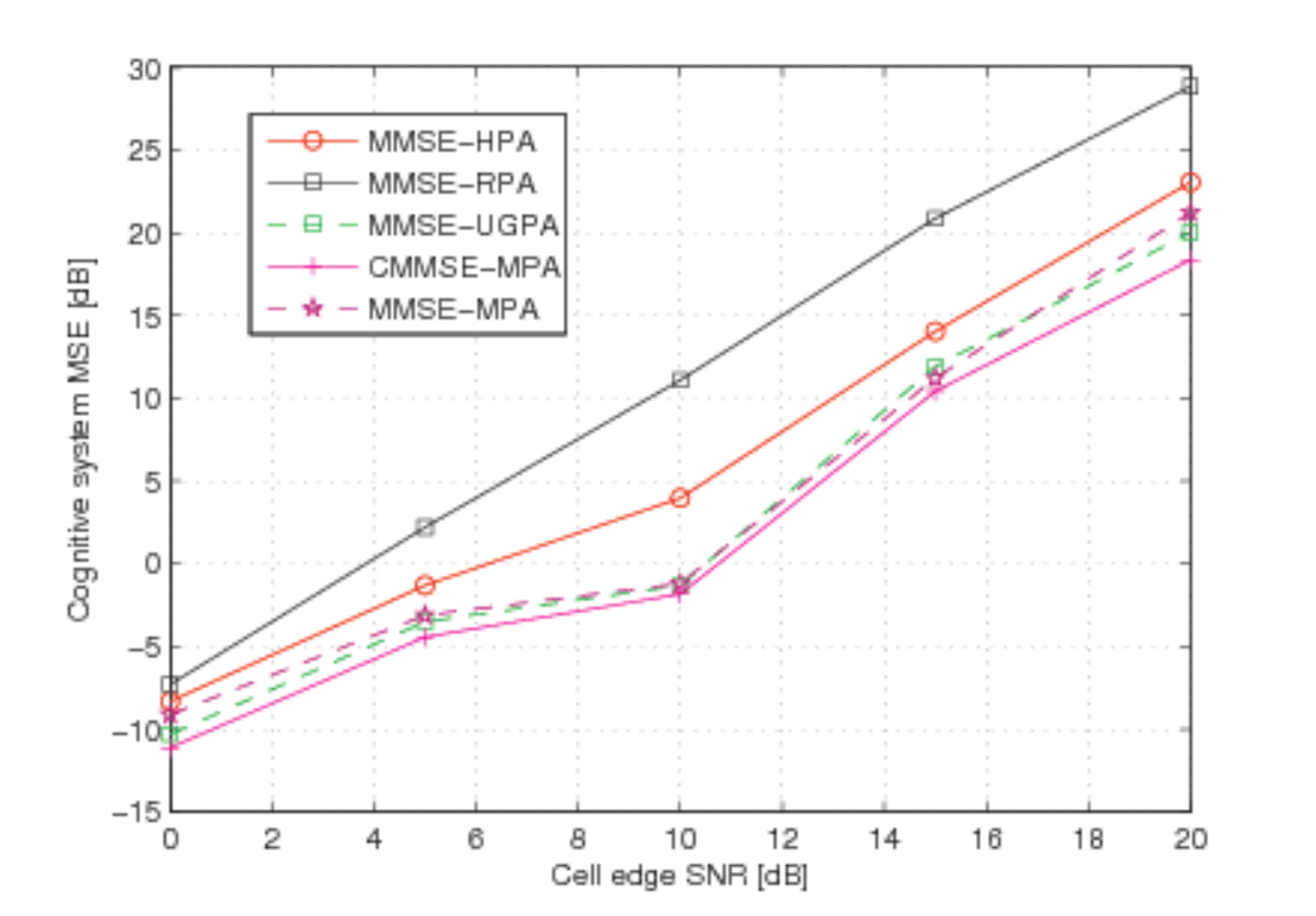}
\caption{\vspace{-0.1cm}\textit{\label{fig1}\small Cognitive system MSE vs cell edge SNR}}
\end{center}
\end{figure}

\begin{figure}[h]
\begin{center}
\includegraphics[scale=0.5]{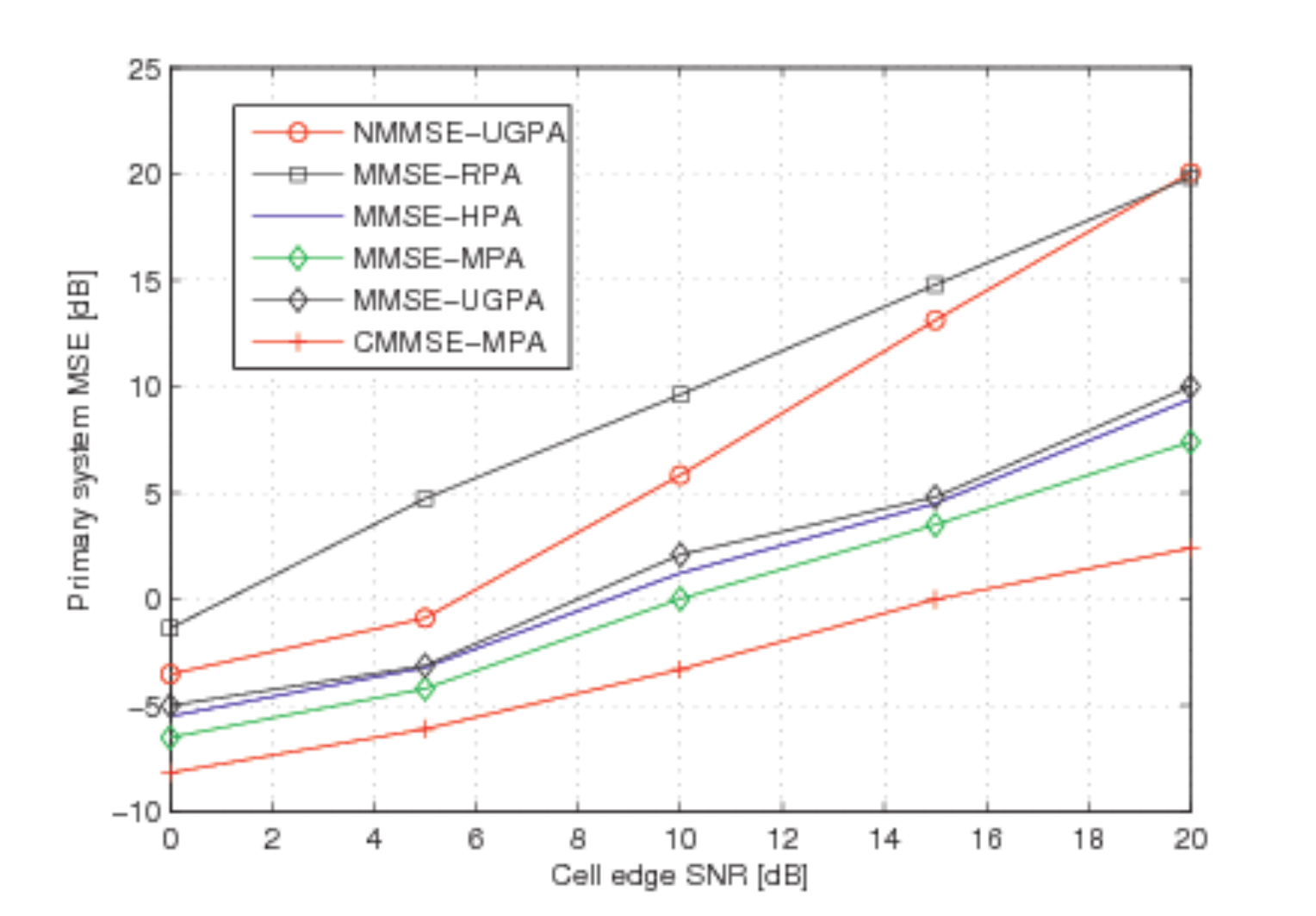}
\caption{\vspace{-0.1cm}\textit{\label{fig2}\small Primary system MSE vs cell edge SNR}}
\end{center}
\end{figure}

\begin{center}
\footnotesize
\begin{table}
\begin{tabular}{|p{2cm}|p{4cm}|p{1cm}|}
\hline
Acronym&Estimation scheme&Equation number\\
\hline
NMMSE&Naive Minimum Mean Square Estimation&(\ref{nmmse})\\
\hline
MMSE& Minimum Mean Square Estimation&(\ref{smmse})\\
\hline
CMMSE&Cognitive Minimum Mean Square Estimation&(\ref{cmmse})\\
\hline
\end{tabular}
\normalsize
\vspace{0.1cm}
\footnotesize
\caption{The list of adopted estimation in simulation}
\end{table}
\end{center}
\begin{center}
\begin{table}
\begin{tabular}{|p{1.5cm}|p{4cm}|p{2cm}|}
\hline
Acronym& Pilot Allocation Scheme&algorithm number\\
\hline
MPA& Mean square error pilot allocation &\textbf{A.1}\\
\hline
HPA&Heuristic pilot allocation&\textbf{A.2}\\
\hline
UGPA& User grouping pilot allocation&\textbf{A.3}\\
\hline
RPA& Random  pilot allocation &\quad\\
\hline
\end{tabular}
\vspace{0.1cm}
\caption{The list of adopted pilot allocation in simulation}
\end{table}
\end{center}
\normalsize
Fig. (\ref{fig1}) depicts the comparison among the different pilot allocation strategies with respect to cell edge SNR. The mean square error
performance for pilot allocation in the cognitive system  is studied, for
nominal reuse factor of $3$. It can be noted that the
random pilot allocation has the worst performance in comparison with the
other techniques. This can be explained by the fact that RPA does not pay any attention about the separability
between the SU and PU and or the other SUs. On the other hand, MSE based
pilot allocation techniques outperform all techniques. User grouping
and heuristic pilot 
allocation achieve a comparable performance with respect to MPA with the
advantage
of reduced complexity.  
 \\
Figure (\ref{fig2}) illustrates the contamination and its impact on
MSE performance of the PU. It can be clearly noted that the CU existence
has a powerful impact on the estimation process. The modified MMSE estimator
with MPA
has a superior performance in the rejection of the interference especially
at SNR, due to its capability of limiting the interference to certain value.
 On other hand, using RPA at the CBS has a very harmful impact on the estimation
 at PBS since it does not take into the account the spatial separability between the
 CUs and the PU. The user grouping PA and heuristic
 PA show less impact on PU in comparison with RPA, which motivates their usefulness
 due their simple implementation.

\section{Conclusion}
In this paper, we discussed the impact of pilot contamination during channel estimation and its influence on primary and cognitive systems. We presented and studied the performance of different MMSE techniques. We proposed  modified
estimation and pilot allocation techniques to tackle the primary-cognitive hierarchy. They enabled enhanced estimation by reducing the overlap in the interfering subspaces and boosting
the separation in the signals through allocating the same pilot to CUs that
has distinct spatial characteristics from PU. Moreover, different pilot allocation techniques
are proposed to enhance the estimation and to reduce the impact of contamination
on the two systems. The performance of introduced algorithms was investigated and compared
to current state of the art techniques. From the simulation results, it can
be concluded that the proposed MMSE estimation  techniques combined with pilot allocations provide considerable gains over the traditional techniques.
\section*{acknowledgment}
This work was supported by the National Research Fund (FNR) of
Luxembourg under the AFR grant (reference 4919957) for the project Smart Resource Allocation Techniques for Satellite Cognitive Radio. 
%\scriptsize


\begin{thebibliography}{1}
\vspace{-0.1cm}
\scriptsize
% \bibitem{federal}
% Federal Communication Commission, Spectrum Policy Task Force, ET
% document no. 02-135, Nov. 2002.\smallskip

\bibitem{goldsmith}
A. Goldsmith, S. A. Jafar, I. Maric, and S. Srinivasa, ``Breaking spectrum
gridlock with cognitive radios: An information theoretic perspective,"
\textit{IEEE}, vol. 97, no. 5, pp. 894 - 914, May 2009.\smallskip

\bibitem{Haykin}
S. Haykin, ``Cognitive Radio: Brain-Empowered Wireless Communications,"
\textit{IEEE Journal on Selected Areas in Communications}, vol. 23, pp. 201-22, Feb. 2005.\smallskip
% 
% \bibitem{lv_o}
% J. Lv, R. Blasco-Serrano, E. Jorswieck, R. Thobaben and A. Kliks,
% ``Optimal Beamforming in MISO Cognitive Channels with Degraded Message Sets," \textit{IEEE Conference on Wireless Communications and Networking (WCNC)}, April 2012.\smallskip
% 
% \bibitem{jorswieck}
% J. Lv, R. Blasco-Serrano, E. Jorswieck and R. Thobaben, ``Linear Precoding
% in MISO Cognitive Channels with Causal Primary Message," \textit{IEEE International
% Symposium on Wireless Communications Systems (ISWCS)}, August 2012.\smallskip

\bibitem{prior}
S.H. Song and K. B. Lataief,``Prior Zero-Forcing for Relaying Primary Signals in
Cognitive Network," \textit{IEEE Global Conference in Telecommunications
(Globecom)},
December, 2011.\smallskip
\bibitem{lee}
K.-J. Lee and I. Lee ``MMSE Based Block Diagonalization for Cognitive Radio MIMO Broadcast Channels,"  \textit{IEEE Transactions on Wireless Communications},
vol. 10, no. 10, pp. 3139 - 3144, October 2011.
\bibitem{shree}
S. K. Sharma, S. Chatzinotas and B. Ottersten,``Spectrum Sensing in Dual Polarized Fading Channels for Cognitive SatComs," \textit{IEEE Global conference on Telecommunications(Globecom)}, December 2012.\smallskip

% \bibitem{gan_t}
% G. Zheng, S. H. Song, K. K. Wong, and B. Ottersten,``Cooperative Cognitive Networks: Optimal, Distributed and Low-Complexity Algorithms," \textit{IEEE Transaction on Signal Processing}, vol. 61, no.11 , pp. 2778 - 2790, June 2013.\smallskip

\bibitem{jose}
J. Jose, A. Ashikhmin, T. L. Marzetta and S. Vishwanath,
``Pilot Contamination and Precoding in Multi-Cell TDD Systems," \textit{IEEE Transactions on Wireless Communications}, vol. 10, no. 8, pp. 2640 - 2651, Aug. 2011.\smallskip
% 
% \bibitem{Jindal}
%  B. Gopalakrishnan and N. Jindal, ``An Analysis of Pilot Contamination on
% Multi-user MIMO Cellular systems with Many Antennas," \textit{IEEE International Workshop on Signal Processing Advances in Wireless
% Communications (SPAWC)}, Jun 2011.\smallskip

\bibitem{gan}
N. Krishnan, R. D. Yates, and  N. B. Mandayam, ``Cellular Systems with Many Antennas: Large System Analysis under Pilot Contamination,"\textit{ Allerton
Conference on Communications, Computing and Control}, October, 2012.\smallskip
\bibitem{MAHA-ICC}
M. Alodeh, S. Chatzinotas and B. Ottersten,``Spatial DCT-Based Least Square Estimation in Multi-antenna Multi-cell Interference Channels,"\textit{ to
appear in IEEE International Conference on Communications (ICC), 2014.}\smallskip
\bibitem{MAHA}
M. Alodeh, S. Chatzinotas and B. Ottersten,``Spatial DCT-Based Channel Estimation in Multi-Antenna Multi-Cell Interference Channels,"\textit{ submitted to IEEE Transactions on Signal Processing, available on Arxiv.}\smallskip
\bibitem{haifan}
H. Yin, D. Gesbert, M. Filippou and Y. Liu, ``A Coordinated Approach to Channel Estimation in Large-scale Multiple-antenna Systems," \textit{IEEE Journal in selected Areas
in Communications}, vol. 31  , no. 2, pp. 264 - 273, February 2013.\smallskip
\bibitem{caire-s}
J. Nam,  J.-Y. Ahn, and G. Caire, ``Joint Spatial division and Multiplexing-The large Scale Array
Regime," \textit{IEEE Transaction
on Information Theory}, vol. 51, no. 5 , pp. 6441 - 6463, October 2013.\smallskip
\bibitem{caire-g}
A. Adhikary, and G. Caire, ``Joint Spatial Division and Multiplexing: Opprtunistic
Beamforming and User Grouping,"\textit{Available on Arxiv}.\smallskip
\bibitem{kammeryer}
A. Scherb and K. Kammeyer, ``Bayesian channel estimation for doubly
correlated MIMO systems," \textit{ IEEE Workshop on Smart Antennas (WSA)},
2007.\smallskip
\bibitem{perz}
P. Zetterberg and B. Ottersten, ``The spectrum efficiency of a base station
antenna array for spatially selective transmission," \textit{IEEE Transactions
on Vehicular Technology}, vol. 44, no. 3, pp. 57-69, Aug. 1995.\smallskip

% \bibitem{cristoff}
% C. Martin, and B. Ottersten, `` Asymptotic eigenvalue distributions and capacity for MIMO channels under correlated fading," \textit{IEEE Transactions on Wireless Communications}, vol. , no. , pp. 1350-1359, 2003.
% \bibitem{marzetta}
% T. L. Marzetta, ``Noncooperative cellular wireless with unlimited numbers of base station antennas," \textit{ IEEE Transactions on Wireless Communications}, vol. 9, no. 11, pp. 3590-3600, Nov. 2010.
% 
% \bibitem{emil}
% E. Bj\"{o}rnson and B. Ottersten,``A Framework for Training-Based Estimation in Arbitrarily Correlated Rician MIMO Channels with Rician Disturbance,"
% \textit{IEEE Transactions on Signal Processing}, vol. 58, no. 3, pp. 1807-1820, March 2010.\smallskip


% \bibitem{mackay}
% S. M. Kay, Fundamentals of Statistical Signal Processing: Estimation Theory. Englewood Cliffs, NJ: Prentice Hall, 1993.\smallskip

\end{thebibliography}
\end{document}